\documentclass[11pt,a4paper]{article}
\pdfoutput=1
\usepackage{jheppub}
\usepackage{mathrsfs}
\usepackage{braket}
\usepackage{amsmath}
\usepackage{verbatim}
\usepackage{amssymb}
\usepackage{bbold}
\usepackage[normalem]{ulem}
\usepackage{inputenc, array}
\usepackage{textcomp}
\usepackage{appendix}
\usepackage{epsfig}
\usepackage{pifont}
\usepackage{multirow}
\usepackage{makecell}
\usepackage{upgreek}
\usepackage{multirow}
\usepackage{jheppub}
\usepackage{enumerate}
\usepackage{relsize}
\usepackage{inputenc}
\usepackage{braket}
\usepackage{graphicx}
\usepackage{chemarrow}
\usepackage{mathtools}
\usepackage{subcaption}
\usepackage[font=small]{caption}
\usepackage{blindtext}
\usepackage{amsmath,lipsum,amssymb,leftindex}
\usepackage{xcolor}
\definecolor{arroworange}{RGB}{245,150,0}
\definecolor{boxyellow}{RGB}{255,248,220}
\usepackage{tikz}
\usetikzlibrary{arrows.meta,positioning}

\newcommand{\n}{\nu}
\newcommand{\m}{\mu}

\newcommand{\xcentcolon}

\usepackage{color}
\newcommand{\e}{\epsilon}
\newcommand{\be}[1]{\begin{equation}\label{#1} }
\newcommand{\ee}{\end{equation}}
\newcommand{\bea}[1]{\begin{eqnarray}\label{#1} }
\newcommand{\eea}{\end{eqnarray}}
\newcommand{\p}{\partial}

\newcommand{\D}{\Delta}

\renewcommand{\a}{\alpha}

\renewcommand{\b}{\beta}
\renewcommand{\t}{\tau}
\newcommand{\s}{\sigma}

\title{Towards a Holographic dual of Carrollian BCFT}

\author[a]{Pronoy Chakraborty,} \author[b]{Ritankar Chatterjee,} \author[c]{and Priyadarshini Pandit.}\author{\\}

\affiliation[a]{Department of Physics, Indian Institute of Technology Ropar, Rupnagar, Punjab 140001, India.\\}

\affiliation[b]{Beijing Institute of Mathematical Sciences and Applications, Beijing 101408, China.\\}

\affiliation[c]{Department of Theoretical Physics, Tata Institute of Fundamental Research, Mumbai, 400005, India.\\}

\emailAdd{pronoy.22phz0009@iitrpr.ac.in, ritankar@bimsa.cn, priyadarshini.pandit@tifr.res.in}

\preprint{}

\abstract{In this work we aim to lay the foundation of flat version of AdS$_3$/BCFT$_2$ correspondence through Carrollian framework. We begin with discussing the recently discovered Boundary Carrollian Conformal Algebra (BCCA),  the symmetry algebra of Carrollian BCFT. We identify this algebra at the null infinity of flat spacetime with appropriate choice of the end-of-the-world (EOW) brane. The properties of this EOW brane have been analysed both intrinsically and by taking flat limit from its AdS counterpart. We also find that the global as well as asymptotic symmetries of flat spacetime restricted by this EOW brane coincide exactly with the symmetries of BCCA. }

\begin{document}

\maketitle
\newpage

\section{Introduction}
Over the past three decades, holographic principle \cite{tHooft:1993dmi,Susskind:1994vu} has been one of the most important framework in understanding quantum gravity, which says that, a theory of quantum gravity in a given spacetime can be equivalently described by a non-gravitational theory living on its lower dimensional boundary. The most successful realization of it is the AdS/CFT correspondence \cite{Maldacena:1997re}, which relates quantum gravity in an asymptotically Anti-de Sitter (AdS) spacetime to a conformal field theory (CFT) defined on its boundary. For three-dimensional (3d) gravity, the dual theory is a 2d CFT whose symmetry algebra consists of two copies of the Virasoro algebra. An earlier hint about AdS$_3$/CFT$_2$ correspondence was provided by the seminal work of Brown and Henneaux \cite{Brown:1986nw}. They showed that the asymptotic symmetry algebra of AdS$_3$ gravity is exactly two copies of Virasoro algebra.

\medskip

CFTs defined on manifold with boundaries, also called boundary conformal field theories (BCFTs), have played an important role in a wide range of physical systems which includes quantum impurity problems and other condensed matter applications \cite{Kondo:1964nea,Affleck:1990zd,Affleck:1990iv,vanWees:1988zz}. Study of 2d BCFTs, whose symmetry algebra reduce to single copy of Virasoro algebra, was first introduced by Cardy \cite{Cardy:1984bb}. They are also essential in the study of open strings and D-branes \cite{Pradisi:1988xd,Polchinski:1987tu,Polchinski:1995mt,Polchinski:1996na,Karch:2000gx,Recknagel:2013uja}. The holographic description of BCFTs was initiated by Takayanagi \cite{Takayanagi:2011zk} and further developed in \cite{Fujita:2011fp}. In this framework, the bulk AdS geometry is restricted by an end-of-the-world (EOW) brane, and the resulting gravitational system provides a holographic dual of a BCFT.

\medskip

Despite the remarkable success of holography to describe gravity in asymptotically AdS spacetimes, its extension to spacetimes with vanishing or positive curvature remains much less understood. In asymptotically flat spacetimes (AFS), two approaches have received considerable attention. The first is celestial holography, in which quantum gravity in AFS is proposed to be dual to CFT defined on a codimension two celestial sphere at null infinity \cite{Strominger:2017zoo,Raclariu:2021zjz,Pasterski:2021rjz}. The second is Carrollian holography, in which the dual theory is a Carrollian conformal field theory (CCFT) living on a codimension one null boundary of the spacetime \cite{Bagchi:2022emh,Donnay:2022aba,Bagchi:2023fbj,Chen:2023naw,Bagchi:2025vri,Ruzziconi:2026bix,Saha:2023hsl}. This approach is based on the asymptotic symmetry and is motivated by the isomorphism between Carrollian conformal algebra (CCA) \cite{Duval:2014lpa,Duval:2014uva,Barnich:2006av} and the asymptotic symmetry algebra of AFS, namely the Bondi-van der Burg-Metzner-Sachs (BMS) algebra \cite{Bondi:1962px,Sachs:1962wk}. One of the interesting features of Carrollian approach is that most of its results can be obtained from the flat-space limits of its AdS/CFT counterparts. More recently, a complementary route to flat-space holography has been proposed by introducing flat EOW branes in AdS and using insights from AdS/BCFT correspondence to reproduce the duality between quantum gravity in AFS and CCFT \cite{Hao:2025ocu}.
\medskip

A Carrollian CFT is defined as a CFT on a Carrollian/null  manifold with CCA as its symmetry algebra. Examples of Carrollian manifolds are null infinity of AFS, event horizon of black holes and light cones. Carrollian structures have also found applications in hydrodynamics \cite{Bagchi:2023ysc,Kolekar:2024cfg} and condensed matter systems \cite{Bidussi:2021nmp,Bagchi:2022eui,Biswas:2025dte}. CCFTs also arise on the worldsheets of tensionless or null strings \cite{Isberg:1993av, Bagchi:2026wcu,Banerjee:2023ekd,Banerjee:2024fbi}\footnote{See also \cite{Cardona:2016ytk,Bagchi:2023cfp, Banerjee:2025bkg} for studies in Carroll strings.}. Compared to the extensive study of Carrollian CFTs, however, Carrollian CFTs with boundaries have remained largely unexplored. Recent work \cite{Bagchi:2024qsb} initiated the study of classical BCFTs in the context of open null strings, leading to the discovery of a new infinite dimensional symmetry algebra, called Boundary Carrollian Conformal Algebra (BCCA)\footnote{The BCCA discovered in \cite{Bagchi:2024qsb} corresponds to the ungauged (bare) open ILST (tensionless string) action. It has recently been found that a fully consistent physical null string possesses an additional Carroll-Weyl gauge symmetry requiring a new projective constraint along with the existing constraints, resulting into an extension of CCA$_2$ as worldsheet gauge symmetry \cite{Sheikh-Jabbari:2026cnj,Sheikh-Jabbari:2026vqh}. In present work, we are considering the algebra which comes from CCA$_2$ and not any of its extended versions.}~\footnote{The supersymmetric version of this algebra has been studied in \cite{Bagchi:2025jgu}}. The representation theory of this algebra has been studied in \cite{Buzaglo:2025nti}, while study of a new representation theory will appear in an upcoming work \cite{upcomingpaper}.
\medskip

The objective of the present work is to take the first steps towards identifying a holographic dual of Carrollian analogue of BCFTs, also referred to as BCCFTs. Here we identify the possible restricted bulk segment of 3d AFS whose asymptotic symmetry algebra at null infinity is BCCA. This setup will provide a natural bulk for a prospective AFS$_3$/BCCFT$_2$ correspondence and may be regarded as the flat-space analogue of the AdS$_3$/BCFT$_2$ construction. An interesting finding of our work is that different choices of boundary on a Carrollian manifold lead to distinct realisations of BCCA, whereas, for relativistic 2d BCFTs, introduction of boundary universally reduces two copies of Virasoro algebras to a single copy. In this work, our main focus will be to BCCA arising from spatial boundary (along which spatial coordinate is fixed).

\medskip

This paper is organised as follows: In Section \eqref{section2}, we briefly review Boundary Carrollian Conformal Algebra. Section \eqref{section3} deals with analysis of end-of-the-world (EOW) brane in flat spacetime, both intrinsically and by taking the flat space limit of its AdS counterpart. In section \eqref{section4}, we derive the global symmetries of flat spacetime restricted by EOW brane and study its corresponding representation theory. We also support these results by taking an appropriate limit from AdS spacetime. In section \eqref{section5}, we proceed further and discuss asymptotic symmetry algebra for 3d flat spacetime restricted by EOW brane. Section \eqref{section6} discusses another possible version of AFS/BCCFT correspondence. Finally in section \eqref{section7}, we conclude with some possible future directions.
\medskip

\section{BCCA: A Brief Recap}\label{section2}
In this section, we briefly review the conformal symmetries on Carrollian manifold and impact of introducing boundaries on these symmetries. The resultant symmetry algebra is called BCCA.

\subsection{Carrollian conformal symmetries}
Carrollian manifolds are non-Riemannian manifolds where the metric is replaced by vector field $v^\m$ and degenerate symmetric rank two tensor $h_{\m\n}$ satisfying $v^\m h_{\m\n}=0$. Flat Carroll manifold enjoys Carroll symmetry which comes from zero light-speed ($c\to 0$) of Poincar\'e symmetry. For flat Carroll manifold where $v^\m=(1,0,\cdots,0)$ and $h
_{\m\n}=\text{diag} (0,1,1,\cdots,1),$ conformal symmetries are generated by killing vectors $\xi^\m$ which satisfy the following equations
\begin{align}
    \mathcal{L}_\xi v^\m=\lambda v^\m,\qquad  \mathcal{L}_\xi h_{\m\n}=-2\lambda h_{\m\n}.
\end{align}
For 2d Carrollian manifold having coordinates ($\s,u$), these killing equations have the following solutions
\begin{align}
    \xi=[f(\phi)+u g'(\phi)]\partial_\s+g(\phi)\partial_u.
\end{align}
On cylindrical manifolds having topology $\mathbb{R}_{null}\times S^1$, with $\phi \sim \phi +2\pi$,  the killing vectors can be Fourier expanded as
\begin{align}\label{BMSCyl}
    L_n =i\,e^{in\phi}(\partial_\phi+inu\,\partial_u),\qquad M_n= i\,e^{in\phi}\partial_u.
\end{align}
These generators famously satisfy the 2d Carrollian Conformal Algebra (CCA$_2$)
\begin{align}\label{BMS}
    [L_{m},L_{n}]&=(m-n)L_{m+n}+\frac{c_L}{12}(m^3-m)\delta_{n+m,0},\nonumber\\
    [L_{m},M_{n}]&=(m-n)M_{m+n}+\frac{c_M}{12}(m^3-m)\delta_{n+m,0},\\
    [M_{m},M_{n}]&=0.\nonumber
\end{align}
This algebra can also be obtained by taking Carroll contraction (shown below in \eqref{oldcon}) on the Virasoro algebra. The Virasoro algebra is given as follows
\begin{align}\label{plusminus}
    [\mathcal{L}_n^\pm, \mathcal{L}_m^\pm] = (n-m) \mathcal{L}^\pm_{n+m} + \frac{c^\pm}{12} \delta_{n+m,0} (n^3-n), \qquad [\mathcal{L}_n^\pm, \mathcal{L}_m^\mp] = 0.
\end{align}
On relativistic cylinder having coordinates ($\s,\t$), the Virasoro generators are given by
\begin{align}\label{cylrel}
    \mathcal{L}^\pm_n=ie^{in\phi^\pm}\partial_{\pm}, \qquad \phi^\pm=\t\pm\phi, \qquad \partial_\pm=\frac{1}{2}(\partial_\t\pm\partial_\phi).
\end{align}
Applying vanishing light-speed limit via ($\t\to\e u,~\phi\to\phi$), along with Carrollian In\"on\"u-Wigner contraction \begin{align}\label{oldcon}
    L_n=\mathcal{L}^+_n-\mathcal{L}^-_{-n},\qquad M_n=\epsilon(\mathcal{L}^+_n+\mathcal{L}^-_{-n}), \qquad \e\to0,
\end{align}
we obtain generators in \eqref{BMSCyl} satisfying CCA$_2$ algebra \eqref{BMS}. The central charges $c_L$ and $c_M$ come from $c^\pm$ as follows
\begin{align}\label{central}
    c_L=c^+-c^-,\qquad c_M=\e(c^++c^-).
\end{align}
\subsection{Introducing boundaries}
\begin{figure}[t]
\centering
    \includegraphics[scale=0.8]{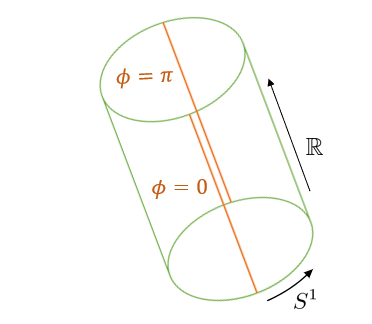}
    \caption{Relativistic cylinder with boundaries}
    \label{cylinder with boundaries}
\end{figure}

Now, let us introduce boundaries on the cylinder at $\phi=0$ and $\pi$ (as shown in Fig.~\ref{cylinder with boundaries}). Here we express Virasoro generators in a different basis (useful when boundaries are introduced), defined below 
\begin{equation}\label{Virnewbas}
\mathbb{L}_n=\mathcal{L}^+_n+\mathcal{L}^-_n, \qquad \widetilde{\mathbb{L}}_n=\mathcal{L}^+_n-\mathcal{L}^-_n
\end{equation}
The above generators $\mathbb{L}_n,\widetilde{\mathbb{L}}_n$ at $\phi=0$ and $\pi$ reduces to 
\begin{align}\label{sigmazero}
    \mathbb{L}_{n}|_{\phi=0}=-e^{in\tau}\partial_{\tau},\,~    \widetilde{\mathbb{L}}_{n}|_{\phi=0}=e^{in\tau}\partial_\phi; \qquad 
    \mathbb{L}_{n}|_{\phi=\pi}=e^{in\tau}\partial_{\tau},\,~ 
    \widetilde{\mathbb{L}}_{n}|_{\phi=\pi}=-e^{in\tau}\partial_\phi.
\end{align}
Here we notice that $\mathbb{L}_n$s has no $\partial_\phi
$ term at $\phi=(0,\pi)$ rendering them as boundary compatible generators (since they leave the boundaries unaltered). However, $\widetilde{\mathbb{L}}_n$s have $\partial_\phi$ term at $\phi=(0,\pi)$, making them boundary incompatible generators.
\medskip

We now apply the same principle on generators of CCA$_2$ \eqref{BMSCyl} written in a different basis. We begin with the following linear combinations of $L_n$s 
\begin{equation}
    \mathcal{O}_n=L_n-L_{-n}, \qquad \mathcal{Q}_n=L_n+L_{-n}
\end{equation}
Using \eqref{BMSCyl} in the above expression, one can see that $\mathcal{O}_n$s are free from any $\partial_\phi$ term at $\phi=(0,\pi)$, while $\mathcal{Q}_n$s aren't, which means $\mathcal{O}_n$ ($\mathcal{Q}_n$) are boundary compatible (incompatible) generators. At first glance, the entire set of $M_n$s appears to be boundary-compatible. However, the limiting perspective (explained below) reveals a very different picture.
\medskip

The generators $\mathcal{O}_n$s and $\mathcal{Q}_ns$ can be obtained from the Carrollian contraction of the Virasoro generators in the new basis \eqref{Virnewbas} as,
\begin{align}\label{BCarCon1}
     \mathcal{O}_{n}=\mathbb{L}_n-\mathbb{L}_{-n}, \qquad \mathcal{Q}_{n}=\widetilde{\mathbb{L}}_n+\widetilde{\mathbb{L}}_{-n}.
\end{align}
This tells us that $\mathcal{O}_n$s come from the limit of relativistic boundary compatible generators, however, $\mathcal{Q}_{n}$s from the limit of discarded generators. Hence, even after expressing $\mathcal{O}_n$s as $\e\to 0$ limit and including subleading terms, we do not get any $\partial_\phi$ term at $\phi=0,\pi$. Now we proceed further to analyse the other set of generators $M_n$s. Different combinations of $M_n$s can be defined as
\begin{equation}\label{BCarcon}
    \mathcal{P}_n=M_n+M_{-n}, \qquad \mathcal{R}_n=M_n-M_{-n}.
\end{equation} 
These generators come from the following limit of Virasoro generators $\mathbb{L}_n, \widetilde{\mathbb{L}}_n$,
\begin{align}\label{BCarCon2}
    \mathcal{P}_n=\e(\mathbb{L}_n+\mathbb{L}_{-n}), \qquad \mathcal{R}_n=\e(\widetilde{\mathbb{L}}_n-\widetilde{\mathbb{L}}_{-n}).
\end{align}
It is clear from the above expression that while $\mathcal{P}_n$s arise from the limit of boundary compatible generators ($\mathbb{L}_n$), $\mathcal{R}_n$s arise from the limit of incompatible generators ($\widetilde{\mathbb{L}}_n$). Consequently, $\mathcal{P}_n$s at $\phi=0,\pi$ contain no $\partial_\phi$ term even in subleading order, but $\mathcal{R}_n$s, when expressed as $\e\to 0$ limit, do have non-vanishing $\partial_\phi$ term at $\phi=0,\pi$ in subleading order, 
\begin{align}\label{khurorkal}
    \mathcal{R}_n\big|_{\phi=0,\pi}=-2in\e^2u ~\cos{(n\phi)}~\p_\phi~\big|_{\phi=0,\pi}~.
\end{align}
Hence, the limiting analysis shows us that $\mathcal{R}_n$s, unlike $\mathcal{P}_n$s, allow small distortions on $\phi=0,\pi$ boundaries, arising as subleading order in $\epsilon$. This renders $\mathcal{R}_n$s boundary incompatible while $\mathcal{P}_n$s remain boundary compatible.
\medskip

Hence, the surviving generators, after introducing boundaries on the null cylinder are $\mathcal{O}_n$s and $\mathcal{P}_n$s 
\begin{equation}\label{unpro7}
\mathcal{O}_{n}=-2~\sin{n\phi}~\partial_{\phi}-2nu~\cos{n\phi}~\partial_{u}, \qquad \mathcal{P}_{n}=2i\cos{n\phi}~\partial_{u}.
\end{equation}
They satisfy the Boundary Carrollian Conformal Algebra (BCCA) given below,
\begin{align}\label{pro952}
    &    [\mathcal{O}_m,\mathcal{O}_{n}]~=~(m-n)\mathcal{O}_{m+n}- (m+n)\mathcal{O}_{m-n},
 \nonumber\\
&   [\mathcal{O}_m,\mathcal{P}_{n}]~=~(m-n)\mathcal{P}_{m+n}+ (m+n)\mathcal{P}_{m-n}+\frac{c_M}{12}(n^3-n)(\delta_{n,m}+\delta_{n,-m}), \\
&  [\mathcal{P}_m,\mathcal{P}_{n}]~=~0. \nonumber
\end{align}
With this we conclude our brief review on BCCA. We now turn our attention to the identification of the possible bulk realisation of BCCA.

\section{Introducing EOW brane in flat space}\label{section3}
It is well known that CCA$_2$ (or BMS$_3$) appears as the asymptotic symmetry algebra at the null infinity of 3d flat spacetime \cite{Barnich:2006av,Bagchi:2010zz}. This fact is the foundation of Carrollian holography, which is a strong candidate of holography of AFS \cite{Barnich:2006av,Bagchi:2010zz,Bagchi:2013qva}. In Carrollian holography proposal, gravity in AFS is dual to the Carroll CFT living on the future null infinity $\mathcal{I}^+$. 
\medskip

In this section, we attempt to find possible bulk dual of the BCCFT living on $\mathcal{I}^+$ which has the topology of Carrollian cylinder $\mathbb{R}_{null}\times S^1$. We would like to look into the possible end-of-the-world (EOW) branes in the flat spacetime, intersection of which with the null infinity gives $\phi=0,\pi$ boundaries on null infinity. 
\medskip

Let us recall that on cylindrical Carrollian manifold the boundaries are located at \{$\phi=0,\pi$\}. We need to find an EOW brane which intersects $\mathcal{I}^+$ at those boundaries. In order to do this we recall the metric of 3d flat spacetime
\begin{equation}\label{3dflat}
\begin{split}
    ds^2=-d&t^2+dr^2+r^2d\phi^2=-(dx^0)^2+(dx^1)^2+(dx^2)^2,\\
    &x^0=t,~~~x^1=r\cos{\phi},~~~x^2=r\sin{\phi}.
    \end{split}
\end{equation}
\begin{figure}[t]
\centering
    \includegraphics[scale=0.65]{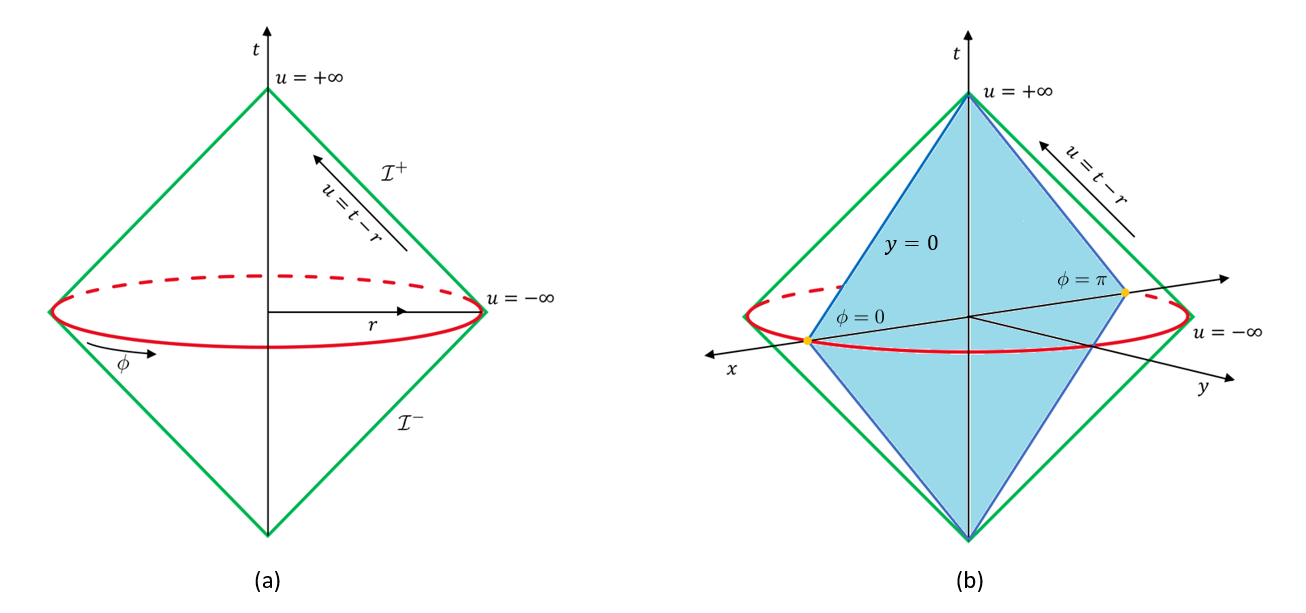}
    \caption{(a) shows Penrose diagram of Minkowski spacetime and (b) shows Minkowski spacetime with EOW brane at $x^2=y=r\sin{\phi}=0$ }
    \label{simple eow brane}
\end{figure}

The coordinate $\phi$ appearing at the null infinity is same as the azimuthal angle appearing in the 3d flat spacetime metric. The simplest EOW brane intersecting with the null infinity at \{$\phi=0,\pi$\} as shown in Fig. \ref{simple eow brane} is given by
\begin{align}\label{EOW1}
    x^2=r\sin{\phi}=0.
\end{align}
One can also consider a more general EOW brane which is $x^2=r\sin{\phi}=-a$ for finite $a$, which also intersects $r\to\infty$ cylinder at \{$\phi=0,\pi$\}.
\subsection{Intrinsic look into the EOW brane}

Let us consider the standard Einstein-Hilbert action without any cosmological constant term with Gibbon's-Hawking-York (GHY) boundary term\footnote{In 3d AFS, for the asymptotic ($r\to\infty$) boundary, a slightly modified GYK boundary term is considered where the coefficient of it is $\frac{1}{16\pi G}$ instead of $\frac{1}{8\pi G}$ \cite{Bagchi:2013lma,Detournay:2014fva}. Using this modification, thermodynamics of 3d flat space cosmology was obtained. However, as shown in \cite{Detournay:2014fva}, variation of the action with constant $r$ hypersurface with large $r$ is of the order $O(1/r)$, i.e. variational principle for that action is well defined only at $r\to\infty$. However, the EOW brane is a boundary placed at finite region where this variation would not vanish and hence the action discussed in \cite{Bagchi:2013lma,Detournay:2014fva} is no longer applicable. The usual GYK term with $\frac{1}{8\pi G}$ coefficient continues to be useful to make the variational principle well defined for the EOW brane we consider here.} along with constant boundary matter Lagrangian
\begin{align}\label{E.H.G.Y.K.}
    I=\frac{1}{16\pi G}\int_{V} d^3x \sqrt{-g}~R+\frac{1}{8\pi G}\int_{Q} d^2y \sqrt{-h}~(K-T).
\end{align}
In the above, $V$ represents 3d bulk region, $Q$ is the 2d EOW brane, $T$ is the uniform localised matter field on $Q$ (also called tension), $K=h^{ab}K_{ab}$ ($a,b=0,1$) where $h_{ab}$ is induced metric on the EOW brane and $K_{ab}$ is the extrinsic curvature of $Q$
\begin{align}
    K_{ab}=e^\a_a e^\b_b\nabla_\a n_\b.  
\end{align}
Variation of action \eqref{E.H.G.Y.K.} w.r.t. metric along with Neumann boundary condition on $Q$ leads us to the following equation of motion on $Q$
\begin{align}
    K_{ab}=(K-T)h_{ab}.
\end{align}
Taking trace of both sides we get $K=2T$. Now, let us consider the EOW brane profile \eqref{EOW1}. This is a flat EOW brane in flat spacetime which means $h_{ab}=\eta_{ab}$ and $\nabla_\a n_\b=0$, i.e. $K_{ab}=0$. That means $K=0$, which in turn, implies that $T=0$ too. Hence, the EOW brane described in \eqref{EOW1} (or more generalised finitely shifted brane) is tensionless (i.e. without any uniform localised matter field). 
\medskip

\subsection{Flat limit of AdS$_3$ EOW branes}\label{Ads3Eow}
\begin{figure}[t]
\centering
    \includegraphics[scale=0.80]{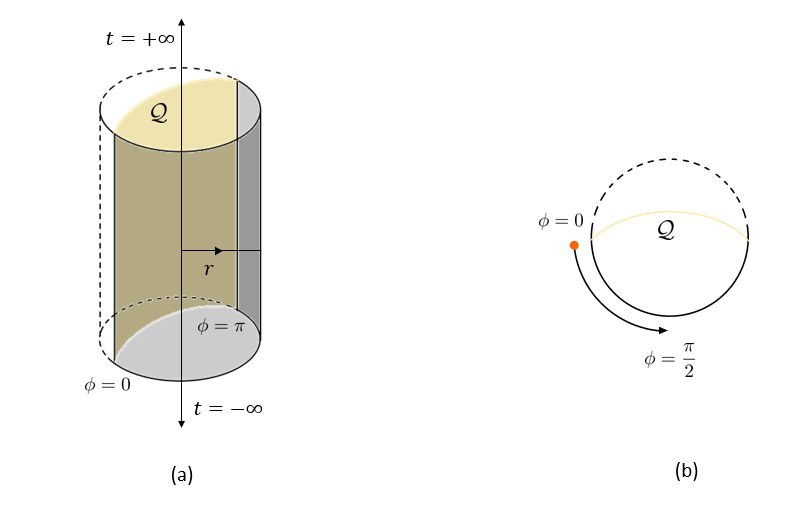}
    \caption{EOW brane in AdS Spacetime (a) along with its constant time slice (b)}
    \label{AdS with EOW brane}
\end{figure}

Since Carrollian holography is also studied as flat-space limit of AdS/CFT correspondence, it is expected that there should be a version of AdS$_3$/BCFT$_2$ correspondence, EOW brane of which, after taking flat-space limit, would lead us to the EOW brane discussed above. The version of AdS$_3$/BCFT$_2$ we should look for in this context is holographic dual of 2d BCFT defined on infinite strip, 
discussed in \cite{Kawamoto:2022etl,Kusuki:2022ozk}, since topologically infinite strip is $\mathbb{R}\times S^1$ restricted by $\phi=0,\pi$ boundaries. Fig. \ref{AdS with EOW brane} illustrates bulk dual of BCFT$_2$ on an infinite strip, where $\mathcal{Q}$ depicts the EOW brane. Fig (a) shows the entire setup and (b) is its constant time slice. Let us begin with the metric of AdS$_3$ spacetime with radius $\ell$
\begin{align}\label{AdS3}
   ds^2&=-f(r)dt^2+f(r)^{-1}dr^2+r^2d\phi^2,\qquad f(r)=\Bigg(1+\frac{r^2}{\ell^2}\Bigg)
\end{align}
 As shown in earlier studies \cite{Kawamoto:2022etl,Kusuki:2022ozk}, the corresponding EOW brane profile in the bulk is $r\sin{\phi}=k$, $k$ being constant. Solving Einstein's gravity with negative cosmological constant and Neumann boundary condition on the EOW brane one can determine $k$ in terms of tension $T$ and AdS radius $\ell$, and get the following EOW brane profile
\begin{align}\label{EOW1AdS}
    r\sin{\phi}=-\frac{T\ell^2}{\sqrt{1-T^2\ell^2}}~.
\end{align}
Taking $\ell\to\infty$ limit on r.h.s of \eqref{EOW1AdS} can make sense iff $T$ is dialed to zero in the order of $\frac{1}{\ell^2}$. Applying tensionless limit $T=\frac{a}{\ell^2}$ on this EOW brane profile one gets the following at $\ell\to\infty$ ($a$ finite)
\begin{align}
    r\sin{\phi}=-a.
\end{align}
\begin{figure}
    \centering    \includegraphics[width=0.85\linewidth]{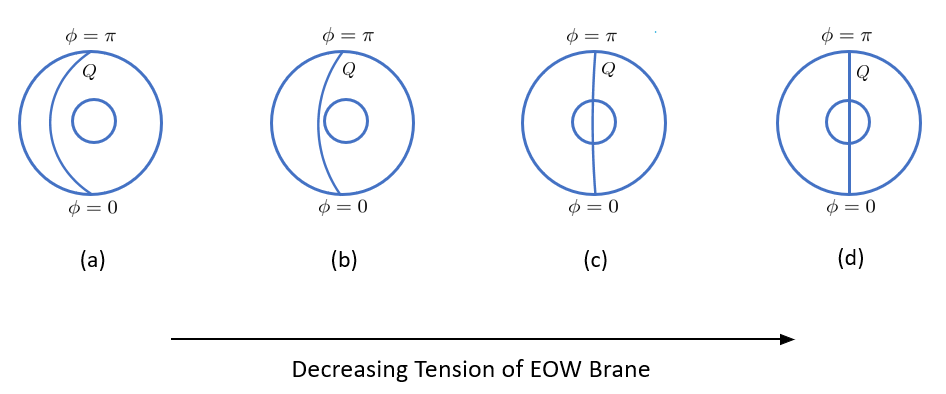}
    \caption{Constant time slices of AdS$_3$ spacetimes with EOW branes with decreasing tension. The circles inside represent the Minkowski diamonds at the center of AdS$_3$. Fig (a) and (b) corresponds to EOW brane with finite tension which does not enter the Minkowski diamond. As the tension decreases, the brane gets closer to the Minkowski diamond and for very small (or vanishing) tension, the EOW brane passes through the Minkowski diamond as flat EOW brane (Fig (c) and (d)).}
    \label{fig:placeholder}
\end{figure}

As can be clearly seen, this EOW brane coincides with the flat EOW brane for Minkowski spacetime discussed before. To physically make sense of this, let us recall that in $\ell\to\infty$ limit, flat space arises at the center of AdS spacetime (where $r\ll \ell$), top view of which is depicted in Fig \ref{fig:placeholder} (details below). 

\begin{itemize}
\item Fig (a) depicts the case where both $T$ and $\ell$ are finite. For this case, the minimum value of $r$ occurs at $\phi=\frac{3\pi}{2}$ where $r>T\ell^2$, i.e. the EOW brane doesn't enter the Minkowski diamond where $r\ll \ell$. 

\item Fig (b) corresponds to decreasing but finite $T$, hence, the EOW brane is closer (still yet to enter) the Minkowski diamond. 

\item Fig (c) corresponds to $T=\frac{a}{\ell^2}$, at $\ell\to\infty$ (to be more precise, $\ell\gg a$), the minima of $r$ is at $r=a\ll \ell$, which means the EOW brane passes through the Minkowski diamond at the center of AdS. Inside the Minkowski diamond it is a flat tensionless EOW brane having a finite shift from the origin as depicted. 

\item Strict tensionless brane ($T=0$) in AdS, passes through $r=0$, which is also the origin of the Minkowski diamond at the center. This correspond to $a=0$ case, which passes through the origin of Minkowski diamond as shown in Fig (d).
\end{itemize}

This completes our discussion on flat EOW brane in flat spacetime, considering both its intrinsic construction and its realisation as the flat limit of AdS$_3$ EOW branes. We now turn to the bulk (flat spacetime restricted by EOW brane) symmetries corresponding to the BCCA. 
\bigskip

\section{Global Symmetries}\label{section4}
In this section we look into the global symmetries of the flat spacetime restricted by EOW brane \eqref{EOW1}. We shall see that this symmetry coincides with the global part of BCCA$_2$ which is a strong indication that the entire BCCA$_2$ would be found as the asymptotic symmetry of this spacetime at null infinity. We further study the flat limit of global symmetries of AdS$_3$ spacetime with $r\sin{\phi}=0$ EOW brane which should reduce to global part of BCCA$_2$. This would be a strong evidence that holography of flat spacetime with $x^2=0$ EOW brane comes from flat limit of AdS/BCFT correspondence for BCFT defined on a strip. 
\subsection{Intrinsic derivation of global symmetries}\label{intringlobal}
It is well known that the global part of CCA$_2$ is isomorphic to $\mathfrak{iso}(2,1)$ i.e, 3d Poincar\'e algebra, the symmetry algebra of 3 dimensional flat spacetime. To see this isomorphism let us begin with the Poincar\'e algebra in 3d.
In 3d flat space-time we have six symmetry generators (3 translations, 2 boosts and 1 rotation generator).
\begin{align}\label{Prorit1}
    P_\mu=i\partial_\mu, \qquad J_{\mu\nu}=i(x_\mu\partial_\nu-x_\nu\partial_\mu):  \qquad (\mu=0,1,2).
\end{align}
The algebra satisfied by the above generators is given by\footnote{We follow the sign convention ($-,+,+,+$).},
\begin{equation}\label{Prorit2}
\begin{split}
      &[J_{\mu\nu}, J_{\rho\sigma}] = i\big(\eta_{\nu\rho}J_{\mu\sigma} - \eta_{\mu\rho}J_{\nu\sigma} + \eta_{\mu\sigma}J_{\nu\rho} - \eta_{\nu\sigma}J_{\mu\rho} \big),\\
&[J_{\mu\nu}, P_\rho] = i\big(\eta_{\rho\nu}P_\mu-\eta_{\mu\rho}P_\nu\big), \quad [P_\mu, P_\nu] = 0.
\end{split}
\end{equation}
  The above 3d Poincar\'e algebra \eqref{Prorit2} is isomorphic to global part of BMS$_3$ algebra. In order to see this, we rewrite this generators \eqref{Prorit1} in a different basis.
 \begin{equation}\label{procare}
 \begin{split}
      &M_0=P_0, \qquad M_{\pm1}=-P_1\mp iP_2,\\
     &L_0=J_{12}, \qquad L_{\pm1}=-J_{02}\pm iJ_{01}.
 \end{split}
 \end{equation}
The algebra satisfied by these new generators reduces to the global part of the BMS$_3$ algebra given in \eqref{BMS}. 
\medskip

In terms of the Bondi coordinates $u,r,\phi$ given by $u=t-r, ~x^1=r\cos{\phi}$ and $x^2=r\sin{\phi}$, 3d Minkowski metric becomes
\begin{align}
    ds^2=-du^2-2dudr+r^2d\phi^2.
\end{align}
The symmetry generators \eqref{procare} in these coordinates become
\begin{equation}
\begin{split}
    &L_0=i\p_\phi,~~~~ L_{\pm1}=ie^{\pm i\phi}\Big[\Big(1+\frac{u}{r}\Big)\p_\phi\pm iu\p_u \mp i(r+u)\p_r\Big],\\
    &M_0=i\p_u,~~~M_{\pm1}=ie^{\pm i\phi}\Big(\p_u\mp\frac{i}{r}\p_\phi-\p_r\Big). 
\end{split}
\end{equation}

The global subalgebra of BCCA$_2$ \eqref{pro952} is isomorphic to $\mathfrak{iso}(1,1)$ algebra
\begin{align}\label{iso(1,1)}
    [\mathcal{O}_{1}, \mathcal{P}_0] = 2\mathcal{P}_1, \qquad  [\mathcal{O}_{1}, \mathcal{P}_1] = 2\mathcal{P}_0, \qquad [\mathcal{P}_0, \mathcal{P}_1] = 0.
\end{align}
This algebra admits a Casimir operator given by
\begin{align}\label{mirkasi}
    \mathcal{M}_\mathcal{B}^2=\mathcal{P}_0^2-\mathcal{P}_1^2.
\end{align}Eigenvalue of $\mathcal{M}_\mathcal{B}^2$ can be used to classify the representation of the global subalgebra \eqref{iso(1,1)} which we shall discuss in the later part of this section.
\medskip

Now we shall see that how in presence of EOW brane at $x^2=0$, the 3d Poincar\'e algebra breaks and gives us $\mathfrak{iso}(1,1)$ in \eqref{iso(1,1)}. Among the Poincar\'e generators given in \eqref{Prorit1}, only those which do not distort $x^2=0$ (i.e. no $\partial_{2}$ at $x^2=0$) will survive. Clearly the following generators cannot survive
\begin{align}
    P_2=i\partial_{2}, \qquad J_{02}=i(x_0\partial_2-x_2\partial_0), \qquad J_{12}=i(x_1\partial_2-x_2\partial_1),
\end{align}
since all of them have $\partial_{2}$ on the plane $x^2=0$. The surviving generators are given by
\begin{align}
    P_0=i\partial_0, \qquad P_1=i\partial_1, \qquad J_{01}=i(x_0\partial_1-x_1\partial_0).
\end{align}
These generators display 2d Poincar\'e algebra denoted by $\mathfrak{iso}(1,1)$. Now, following the redefinition in \eqref{procare} one can easily make the following identifications
\begin{align}\label{iso(1,1)2}
   \mathcal{O}_1=(L_1-L_{-1})=2iJ_{01},\qquad \mathcal{P}_1=M_1+M_{-1}=-2P_1,\qquad \mathcal{P}_0=2M_0=2P_0.
\end{align}
Rewritten in this basis \eqref{iso(1,1)2}, $\mathfrak{iso}(1,1)$ generators satisfy \eqref{iso(1,1)}.
In Bondi coordinates, they take the following form
\begin{equation}\label{BCCABondi}
    \begin{split}
        \mathcal{O}_1=-2\Big(1+\frac{u}{r}\Big)&\sin{\phi}\p_\phi-2u\cos{\phi}\p_
    u+2(u+r)\cos{\phi}\p_r,~~\mathcal{P}_0=2i\p_u,\\
    &\mathcal{P}_1=2i\cos{\phi}\p_u+\frac{2i}{r}\sin{\phi}\p_\phi-2i\cos{\phi}\p_r.
    \end{split}
\end{equation}

\subsection{Flat limit of AdS$_3$ global symmetries}
In this section, we consider  AdS$_3$ spacetime restricted by the EOW brane \eqref{EOW1AdS} with $T=0$, i.e. $r\sin{\phi}=0$. We shall see how the global symmetry algebra of AdS$_3$ ($\mathfrak{so}
(2,2)$) is broken to $\mathfrak{so}(2,1)$ (or $\mathfrak{sl}(2,R)$) algebra. By taking appropriate In\"on\"u-Wigner contraction on the $\mathfrak{so}(2,1)$ algebra we obtain $\mathfrak{iso}(1,1)$ algebra \eqref{iso(1,1)}.
\medskip

Let us first note that the AdS$_3$ metric as given in \eqref{AdS3} would reduce to Minkowski spacetime in terms of $t,r,\phi$ coordinates. We choose to work in the following coordinates  (following \cite{Hao:2025btl})
\begin{align}\label{ucoordinate}
    u=t-\ell\tan^{-1}{\frac{r}{\ell}},~r,~\phi.
\end{align}
The AdS$_3$ metric in this coordinate becomes
\begin{align}\label{AdS3modified}
    ds^2=-\Big(1+\frac{r^2}{\ell^2}\Big)du^2-2dudr+r^2d\phi^2.
\end{align}
The advantage of working in these coordinates is that at the Minkowski diamond inside the AdS$_3$ spacetime ($\ell\gg r$), these coordinates naturally reduce to Bondi coordinates. We now look into the $\mathfrak{so}(2,2)$ symmetry generators of AdS$_3$,
\begin{align}\label{adsglobal3}
    \mathcal{L}^+_{\pm1}&=\pm\frac{1}{2}e^{\pm i(\frac{u}{\ell}+\tan^{-1}{\frac{r}{\ell}}+\phi)}\sqrt{r^2+\ell^2}\Bigg[\frac{\ell}{r\mp i\ell}\p_u+\frac{1}{r}\p_\phi\mp i\p_r \Bigg],\nonumber\\
    \mathcal{L}^\pm_0 &=\frac{i}{2}(\ell\p_u\pm\p_\phi),\\
    \mathcal{L}^-_{\pm1}&=\pm\frac{1}{2}e^{\pm i(\frac{u}{\ell}+\tan^{-1}{\frac{r}{\ell}}-\phi)}\sqrt{r^2+\ell^2}\Bigg[\frac{\ell}{r\mp i\ell}\p_u-\frac{1}{r}\p_\phi\mp i\p_r \Bigg].\nonumber
\end{align}
Under flat limit ($\ell\to\infty$) (which mirrors the Carrollian limit ($\e\to0$) discussed earlier in section \ref{section2}), these generators reduce to global BMS$_3$ generators through the In\"on\"u-Wigner contraction \eqref{oldcon} for $n=0,\pm1$.\footnote{Here it is important to note that the $\mathfrak{so}(2,2)$ generators in \eqref{adsglobal3} are slightly modified version of the $\mathfrak{so}(2,2)$ generators in \cite{Hao:2025btl} without altering the commutators. Here $-i~(+i)$ has been multiplied to the form of $\mathcal{L}_{1}~(\mathcal{L}_{-1})$ given in \cite{Hao:2025btl}. Consequently, the contraction we use here remains same as the usual Carrollian contraction \eqref{oldcon} instead of the contraction introduced in \cite{Hao:2025btl}. Our rationale behind working with \eqref{adsglobal3} is this: at $r\gg \ell$ the generators in \eqref{adsglobal3} take the following exact form (without any overall factor)
\begin{align}
    \mathcal{L}^\pm_n=ie^{inx^\pm}\Big(\p_\pm-\frac{i}{2}nr\p_r\Big), \qquad x^\pm=\frac{u}{\ell}\pm\phi,\qquad n=0,\pm1.
\end{align}
One can see that on the $r\to\infty$ cylinder (which is a constant $r$ hypersurface) these generators exactly coincide with Virasoro generators on cylinder given in \eqref{cylrel} with $\t$ replaced by $\frac{u}{\ell}$.
}
\medskip

Now, let us introduce an EOW brane at $r\sin{\phi}=0$. Since $r$ can take any real positive value, for such hypersurface, $\sin{\phi}=0$, i.e. $\phi=0,\pi$. That means introducing such boundary in the bulk would remove any generator having $\p_\phi$ term at $\phi=0,\pi$, leaving us with the following generators.
\begin{align}\label{BADSSymm}
    \mathbb{L}_1&=\mathcal{L}^+_1+\mathcal{L}^-_1=e^{i(\frac{u}{\ell}+\tan^{-1}{\frac{r}{\ell}})}\sqrt{r^2+\ell^2}\Bigg[\frac{\ell\cos{\phi}}{r-i\ell}\p_u+\frac{i}{r}\sin{\phi}\p_\phi-i\cos{\phi}\p_r\Bigg],\nonumber\\
   \mathbb{L}_0&=\mathcal{L}^+_0+\mathcal{L}^-_0= i \ell\p_u \\ \mathbb{L}_{-1}&=\mathcal{L}^+_{-1}+\mathcal{L}^-_{-1}=e^{-i(\frac{u}{\ell}+\tan^{-1}{\frac{r}{\ell}})}\sqrt{r^2+\ell^2}\Bigg[\frac{\ell\cos{\phi}}{r+i\ell}\p_u+\frac{i}{r}\sin{\phi}\p_\phi+i\cos{\phi}\p_r\Bigg]. \nonumber
\end{align}
These generators satisfy $\mathfrak{so}(2,1)$ algebra.
In flat space limit, we take the In\"on\"u-Wigner contraction given in \eqref{BCarCon1} and \eqref{BCarCon2} on generators \eqref{BADSSymm} to obtain boundary Carrollian generators $\mathcal{O}_n$ and $\mathcal{P}_n$ for $n=0,\pm1$. 
\begin{align}\label{globccacon}
    \mathcal{O}_1=(\mathbb{L}_1-\mathbb{L}_{-1}),\qquad \mathcal{P}_1=\frac{1}{\ell}(\mathbb{L}_1+\mathbb{L}_{-1}), \qquad \mathcal{P}_0=\frac{2}{\ell}\mathbb{L}_0.
\end{align}
At $\ell\to\infty$ limit they reduce to global BCCA generators \eqref{BCCABondi}.

\begin{figure}
    \centering
\begin{tikzpicture}[
    box/.style={
        draw=black,
        fill=boxyellow,
        rounded corners=7pt,
        minimum width=3cm,
        minimum height=1.3cm,
        align=center,
        inner sep=6pt
    },
    flow/.style={
        -{Stealth[length=4mm,width=3mm]},
        draw=arroworange,
        line width=1.1pt
    },
    arrowlabel/.style={
        font=\small,
        fill=white,
        inner sep=2pt
    },
    node distance=2.5cm and 3.5cm
]

\node[box] (ads) {
    $\mathrm{AdS}_{d+1}$\\[-1mm]
    {\small $\mathrm{SO}(2,d)$}
};

\node[box, right=of ads] (bcft) {
    $\mathrm{BCFT}_{d}$\\[-1mm]
    {\small $\mathrm{SO}(2,d-1)$}
};

\node[box, below=of ads] (flat) {
    $\mathrm{Flat}_{d+1}$\\[-1mm]
    {\small $\mathrm{ISO}(1,d)$}
};

\node[box, below=of bcft] (bccft) {
    $\mathrm{BCCFT}_{d}$\\[-1mm]
    {\small $\mathrm{ISO}(1,d-1)$}
};

\draw[flow]
    (ads.east) -- node[arrowlabel, above] {Boundary} (bcft.west);

\draw[flow]
    (flat.east) -- node[arrowlabel, below] {Boundary} (bccft.west);

\draw[flow]
    (ads.south) -- node[arrowlabel, left, align=center]
    {Flat\\[-1mm] limit} (flat.north);

\draw[flow]
    (bcft.south) -- node[arrowlabel, left, align=center]
    {Carroll\\[-1mm] limit} (bccft.north);

\end{tikzpicture}

    \caption{Schematic diagram of global symmetry breaking in presence of EOW branes.}

    \label{symmetry-square}
\end{figure}
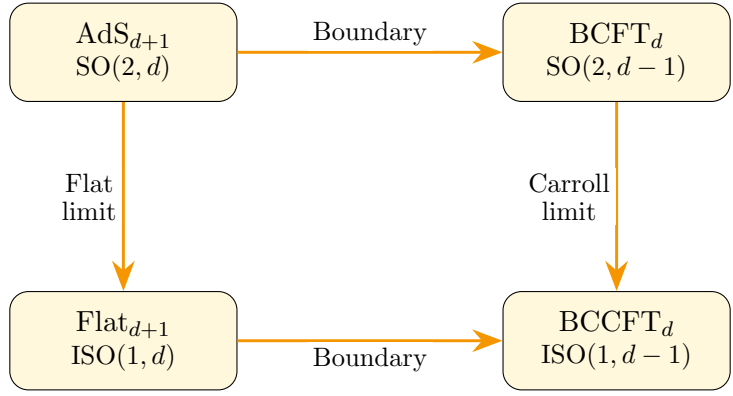

\subsection*{Casimir operator}
Now we proceed further to obtain the Casimir operator for $\mathfrak{iso}(1,1)$ derived in \eqref{mirkasi} from the flat limit ($\ell \to \infty$) of $\mathfrak{so}(2,1)$ algebra. We begin with the Casimir operator of $\mathfrak{so}(2,1)$ given by
\begin{align}\label{BoldC}
        \mathbb{C}=\mathbb{L}^2_0-\frac{1}{2}(\mathbb{L}_{1}\mathbb{L}_{-1}+\mathbb{L}_{-1}\mathbb{L}_{1}).
\end{align}
Inverting \eqref{globccacon} and applying it on \eqref{BoldC} along with $\ell\to\infty$ limit we obtain
\begin{subequations}\label{limCasimir}
\begin{align}\label{limCasimira}
    \mathbb{C}=\frac{\ell^2 \mathcal{P}_0^2}{4}-\frac{1}{4}&\left(\ell^2\mathcal{P}_1-\mathcal{O}_1^2\right) = \frac{\ell^2}{4}\left(\mathcal{P}_0^2-\mathcal{P}_1^2\right)\\ \label{limCasimirb}
    \implies ~&\frac{4 \mathbb{C}}{\ell^2}=  \left(\mathcal{P}_0^2-\mathcal{P}_1^2\right).
\end{align}
\end{subequations}
In the r.h.s of \eqref{limCasimira}, we applied ($\ell \to \infty$) limit where $\mathcal{O}(\ell^2)$ terms dominate. We identify the resultant operator in \eqref{limCasimirb} as the Casimir operator for global BCCA ($\mathcal{M}_B^2$) given in \eqref{mirkasi}. We thus see that applying the flatspace limit ($\ell \to \infty$) on the Casimir operator of $\mathfrak{so}(2,1)$ in \eqref{BoldC}, we obtain Casimir operator for 2d Poincar\'e ($\mathfrak{iso}(1,1)$) or global BCCA.
\medskip

One can check \eqref{limCasimir} more explicitly by applying \eqref{BADSSymm} on \eqref{BoldC} and taking the explicit flat space limit on the resultant expression. At $\ell\to\infty$,  the expression of $\frac{4 \mathbb{C}}{\ell^2}$ reduces to $\mathcal{M}^2_B=\mathcal{P}_0^2-\mathcal{P}_1^2$, where $\mathcal{P}_{0,1}$ are given in \eqref{BCCABondi}.

\subsection*{Some comments on general dimensions}
It is now well known that the global symmetry algebra of AdS$_{d+1}$ is isomorphic to $\mathfrak{so}(2,d)$, which in turn, is isomorphic to conformal symmetry group in $d$ dimensional spacetime\footnote{For $d=2$ the symmetry group becomes infinite dimensional and $\mathfrak{so}(2,2)$ is global subalgebra.}. Introducing a $d-1$ dimensional boundary on the CFT defined on $d$ dimensional spacetime causes breaking of $\mathfrak{so}(2,d)$ symmetry into $\mathfrak{so}(2,d-1)$ symmetry. This results into $d$-dimensional BCFT. In $d+1$-dimensional flat spacetime, the global symmetry algebra is $\mathfrak{iso}(1,d)$ which can also be retrieved from flat limit of $\mathfrak{so}(2,d)$. In presence of boundary, this symmetry break into $\mathfrak{iso}(1,d-1)$, which mirrors the $\mathfrak{so}(2,d)$ to $\mathfrak{so}(2,d-1)$ symmetry breaking in case of AdS/BCFT correspondence. This is illustrated in Fig \ref{symmetry-square}.

\subsection{Representation theory of global symmetries}\label{repglobal}
In this subsection we study the representation theory of $\mathfrak{iso}(1,1)$ and show how the same representation emerges from the Carrollian limit of representation of $\mathfrak{so}(2,1)$\footnote{The representation theory for the entire BCCA$_2$ will appear in an upcoming work \cite{upcomingpaper}.}.

\subsection*{Representation of $\mathfrak{iso}(1,1)$ from induced representation of $\mathfrak{iso}(2,1)$}
Earlier in this section, we showed that, in presence of a $x^2=0$ boundary, the global BMS$_3$ algebra (which is isomorphic to $\mathfrak{iso}(2,1)$) reduces to $\mathfrak{iso}(1,1)$ algebra. We now take this analysis further and obtain a representation theory of $\mathfrak{iso}(1,1)$ from a known representation of $\mathfrak{iso}(2,1)$. Let us look into the induced representation of  $\mathfrak{iso}(2,1)$ \cite{Barnich:2014kra}. In induced representation the module is created on rest frame state $\ket{M,s}$ having momentum $(M,0,0)$ in 3d Minkowski spacetime \cite{Campoleoni:2016vsh}. Such a state can be expressed in terms of global CCA$_2$ algebra generators as 
\begin{align}\label{mamarbarirabdar}
    L_0\ket{M,s}=s\ket{M,s},\quad M_0\ket{M,s}=M\ket{M,s},\quad M_{n}\ket{M,s}=0,\qquad n=\pm 1.
\end{align}
Induced representation of the entire CCA$_2$ algebra is just a generalisation of \eqref{mamarbarirabdar} with generic $n$. Now, let us rewrite \eqref{mamarbarirabdar} in terms of \{$\mathcal{O},\mathcal{P},\mathcal{Q},\mathcal{R}$\}.
\begin{align}
    L_0\ket{M,s}&=\frac{1}{2}(\mathcal{O}_0+\mathcal{Q}_0)\ket{M,s}=s\ket{M,s}\nonumber\\
    M_0\ket{M,s}&=\frac{1}{2}(\mathcal{P}_0+\mathcal{R}_0)\ket{M,s}=M\ket{M,s}\\
    M_n\ket{M,s}&=\frac{1}{2}(\mathcal{P}_n+\mathcal{R}_n)\ket{M,s}=0,\qquad n=\pm 1.\nonumber
\end{align}
Introducing a boundary in the system means discarding all incompatible generators. At the representation level, this translates into an additional restriction on the states: all the incompatible generators annihilate the states
\begin{align}
   \mathcal{Q}_n\ket{M,s}=\mathcal{R}_n\ket{M,s}=0,\qquad n=0,\pm1.
\end{align}
Since $\mathcal{O}_n=-\mathcal{O}_{-n}$, $\mathcal{O}_0$ vanishes and hence imposing $\mathcal{Q}_0\ket{M,s}=0$ would essentially mean $s=0$. Similarly, imposing $\mathcal{R}_0\ket{M,0}=0$ on the state results in $\mathcal{P}_0\ket{M,0}=2M\ket{M,0}$. Now, after imposing $\mathcal{R}_n\ket{M,0}=0$, the state $\ket{M,0}$ can be rewritten as $\ket{\mathcal{M}}$ ($\mathcal{M}=2M$) 
\begin{align}
    \mathcal{P}_0\ket{\mathcal{M}}=\mathcal{M}\ket{\mathcal{M}},\qquad \mathcal{P}_n\ket{\mathcal{M}}=0 \qquad \forall n=\pm 1.
\end{align}
Now, let us look into the descendant states of the 3d Poincar\'e modules, which give us the boosted states in three dimensions
\begin{align}
    \ket{\ell,m}=(L_1)^\ell(L_{-1})^m\ket{\mathcal{M}}
\end{align}
We now rewrite them in terms of \{$\mathcal{O},\mathcal{R}$\}
\begin{align}
    \ket{\ell,m}=(\mathcal{O}_1+\mathcal{Q}_1)^\ell(\mathcal{O}_{-1}+\mathcal{Q}_{-1})^m\ket{\mathcal{M}}.
\end{align}
In order to understand the fate of $\ket{\ell,m}$ in the presence of a boundary, we make the following observation 
\begin{align}\label{lm}
 (\mathcal{O}_\ell+\mathcal{Q}_\ell)(\mathcal{O}_m+\mathcal{Q}_m)
 &=\mathcal{O}_\ell\mathcal{O}_m+[\mathcal{Q}_\ell,\mathcal{O}_m]+\mathcal{O}_m\mathcal{Q}_\ell+\mathcal{O}_\ell\mathcal{Q}_m+\mathcal{Q}_\ell\mathcal{Q}_m.
\end{align}
Using
\begin{equation}
[\mathcal{Q}_\ell,\mathcal{O}_m]=(\ell-m)\mathcal{Q}_{\ell+m}-(\ell+m)\mathcal{Q}_{\ell-m},
\end{equation}
in \eqref{lm} and acting \eqref{lm} on $\ket{\mathcal{M}}$ while imposing that all $\mathcal{Q}$s annihilate $\ket{\mathcal{M}}$ results into the following state
\begin{align}
    \mathcal{O}_l\mathcal{O}_m\ket{\mathcal{M}}.
\end{align}
For the state $\ket{\ell,m}$, we need to apply this method repeatedly, and we end up with the following
\begin{align}
    \ket{\ell,m}=(\mathcal{O}_1)^\ell (\mathcal{O}_{-1})^m\ket{\mathcal{M}}=(-1)^m(\mathcal{O}_1)^{\ell+m}\ket{\mathcal{M}}.
\end{align}
Here we have used the fact that $\mathcal{O}_{-m}=-\mathcal{O}_m$. Hence we end up with the following module for $\mathfrak{iso}(1,1)$ 
\begin{equation}\label{iso11module}
    \begin{split}
\mathcal{P}_0\ket{\mathcal{M}}=\mathcal{M}&\ket{\mathcal{M}},\qquad \mathcal{P}_1\ket{\mathcal{M}}=0,\\
    \ket{\ell}&=(\mathcal{O}_1)^\ell\ket{\mathcal{M}}.
    \end{split}
\end{equation}
We recall from \eqref{mirkasi} that this algebra supports only one quadratic Casimir operator, namely $\mathcal{M}_\mathcal{B}^2=\mathcal{P}_0^2-\mathcal{P}_1^2.$ Eigenvalue of this operator is same for the entire module \eqref{iso11module} which is $\mathcal{M}^2$, i.e. this module forms an irreducible representation of $\mathfrak{iso}(1,1)$. This eigenvalue classifies the entire module. Below we briefly discuss the physical interpretation of this representation of $\mathfrak{iso}(1,1)$.
\subsection*{Physical interpretation}
In 2d flat spacetime, this representation has a straightforward physical interpretation which works the same way as the induced representation of $\mathfrak{iso}(2,1)$. The state $\ket{\mathcal{M}}$ describes a particle of mass $M=\mathcal{M}/2$\footnote{Since $\mathcal{P}_0=2P_0$ and mass of the rest frame particle ($M$) is eigenvalue of $\mathcal{P}_0
$.} at its rest frame. $\mathcal{O}_1$ is the boost generator, and a finitely boosted state is given by 
\begin{subequations}
    \begin{align}
&~~~\ket{\mathcal{M},\eta}=e^{\frac{\eta}{2}\mathcal{O}_1}\ket{\mathcal{M}},\\
\mathcal{P}_0\ket{\mathcal{M},\eta}=\mathcal{M}\cosh{\eta}&\ket{\mathcal{M},\eta},\qquad \mathcal{P}_1\ket{\mathcal{M},\eta}=\mathcal{M}\sinh{\eta}\ket{\mathcal{M},\eta}.
\end{align}
\end{subequations}
The state $\ket{\mathcal{M},\eta}$ belongs to the $\mathfrak{iso}(1,1)$ module described in \eqref{iso11module} since it comes from linear combination of states $\ket{\ell}$ described in \eqref{iso11module}. Absence of spin $s$ in this module is expected since spin comes from spatial rotation, which is absent in 2d flat spacetime (having only one spatial coordinate). The  Casimir operator $\mathcal{M}_\mathcal{B}^2=\mathcal{P}_0^2-\mathcal{P}_1^2$ corresponds to the mass squared of the particle. The fact that it has same eigenvalue for the entire module physically means that the mass-squared of the particle is boost invariant.
\medskip

Interpreting this representation in 3d flat spacetime with $x^2=0$ EOW brane is more complicated. The most likely interpretation would be massive particle whose movement is restricted to a constant $x^2$ hypersurface (since such hypersurfaces only allow movements along $x^0$ and $x^1$ directions).
\medskip

This completes our construction of representations of $\mathfrak{iso}(1,1)$ from induced representations of $\mathfrak{iso}(2,1)$. This discussion of representation theory can be generalised to the entire BCCA. We now show that the same representations can also be obtained by taking the Carrollian/flat limit of representations of a single copy of $\mathfrak{so}(2,1)$.
\subsection{Representation of $\mathfrak{iso}(1,1)$ from limit of $\mathfrak{so}(2,1)$ representation}
The highest weight representation of $\mathfrak{so}(2,1)$ is constructed on the primary states defined below 
\begin{align}\label{H.W.R.}
    \mathbb{L}_0\ket{\Delta}=\Delta\ket{\Delta}, \qquad \mathbb{L}_1\ket{\D}=0.
    \end{align}
Inverting the contraction for $\mathbb{L}_1$ given in \eqref{globccacon} and applying on \eqref{H.W.R.} one gets the following
\begin{align}
    \Bigg(\ell\mathcal{P}_{1}+\mathcal{O}_{1}\Bigg)\ket{\Delta}=0.
\end{align}
At $\ell\to \infty$, where $\ket{\D}\to\ket{\mathcal{M}}$, we have
\begin{align}
    \mathcal{P}_{1}\ket{\mathcal{M}}=0.
\end{align}
From the contraction \eqref{globccacon}, we also have $\mathcal{P}_{0}=\frac{2\mathbb{L}_{0}}{\ell}$. Along with the fact that $\mathbb{L}_0\ket{\D}=\D\ket{\D}$, this leads us to 
\begin{align}\label{weightlim}
   \mathcal{P}_{0}\ket{\Delta}=\frac{2\mathbb{L}_{0}}{\ell}\ket{\Delta}=\frac{2\Delta}{\ell}\ket{\Delta},\qquad \implies \mathcal{M}=\frac{2\Delta}{\ell}.
\end{align}
This confirms that the primary state of the highest weight representation of $\mathfrak{so}(2,1)$ algebra in the flatspace limit ($\ell \to 0$) does reduce to the rest frame state of $\mathfrak{iso}(1,1)$ (below we can see this more concretely).
\medskip

In the context of AdS/BCFT correspondence the highest weight state $\ket{\Delta}$ defined on the BCFT side (which corresponds to a primary state of scaling dimension $\Delta$) is holographically dual to lowest energy state of a massive particle in AdS. For massive scalar field of mass $M$ in AdS$_3$, the relation between $M$ and $\Delta$ is \cite{Aharony:2003qf,Karch:2017wgy}
\begin{align}
    \Delta=1+\sqrt{1+M^2\ell^2}.
\end{align}
In the limit $\ell\to\infty$ this leads to $\Delta\to M\ell=\mathcal{M}\ell/2$, which coincides with \eqref{weightlim}. 
\medskip

Recently, scalar excitations in 3d flat spacetime have been found to be holographically dual to CCFT states belonging to the induced representation of CCA$_2$ \cite{Hao:2025btl}. It has also been found that this duality coincides with flat limit of AdS$_3$/CFT$_2$ correspondence. The results discussed in this section strongly indicate that in case of 3d flat spacetime with EOW brane, the bulk excitations would be dual to BCCFT$_2$ states constructed on new representation discussed in this section. Also we should be able to reproduce this duality from flat limit of AdS$_3$/BCFT$_2$ correspondence for BCFT$_2$ defined on a strip.
\medskip

With this we conclude our analysis of the global symmetries of flat spacetime in the presence of an EOW brane ($x^2=0$). We have derived both the symmetry algebra and its representations intrinsically, as well as from flat space limit of their AdS counterpart. In the next section, we extend this analysis to the corresponding asymptotic symmetries.

\section{Asymptotic symmetries}\label{section5}
In this section we show how BCCA$_2$ appears as asymptotic symmetry algebra of 3d flat spacetime restricted by $x^2=0$ (or, more generally, $x^2=-a$) plane. We also look into the asymptotic symmetries of AdS$_3$ spacetime in presence of EOW branes described in \eqref{EOW1AdS} and show how they reduce to BCCA$_2$ in the flat space limit. This result would become foundation for holography of AFS in presence of EOW brane.

\subsection{Intrinsic analysis}
We begin with an intrinsic analysis of the asymptotic symmetries of 3d flat spacetime. After briefly revisiting the symmetries of 3d flat spacetime at null infinity, we impose the restrictions arising from the presence of EOW brane. Our focus here is on future null infinity $\mathcal{I}^+$.
\medskip

We have seen in \eqref{procare} how the translation and Lorentz transformation generators can be rewritten in a new basis to obtain global CCA$_2$ generators. Rewriting the 3d Minkowski metric in terms of retarded time $u=t-r$, radial coordinate $r$, and azimuthal angle $\phi$ gives 
\begin{align}
    ds^2=-du^2-2du dr+r^2d\phi^2.
\end{align}
Killing vectors of this spacetime (which are also the generators of Poincar\'e algebra) in terms of these retarded time $\t=t-r$, radial coordinate $r$ and azimuthal angle $\phi$ take the following form
  \begin{align}\label{girgitia}
   \xi=\Big(X(\phi)-\frac{u}{r}X''(\phi)-\frac{Y'(\phi)}{r}\Big)\partial_\phi+&\Big(Y(\phi)+u X'(\phi)\Big)\partial_u \nonumber\\&-\Big(rX'(\phi)-u X'''(\phi)-Y''(\phi)\Big)\partial_r.
\end{align}  
In the above $X(\phi)$ and $Y(\phi)$ are periodic functions of $\phi$ and they are spanned by
$e^{im\phi},~~m=0,\pm1.$
For $Y(\phi)=0, X(\phi)=e^{im\phi}$ ($m=0,\pm1$), $\xi$ gives the Lorentz transformation generators and for $X(\phi)=0, Y(\phi)=e^{im\phi}$ ($m=0,\pm1$) $\xi$ as given in \eqref{girgitia} gives the translation generators.
\medskip

The future null infinity $\mathcal{I}^+$ appears at $r\to\infty$ which consists of coordinates $u,\phi$. As shown in \cite{Barnich:2006av} this symmetry is enhanced to infinite dimensional symmetry algebra where $X(\phi)$ and $Y(\phi)$ are arbitrary periodic functions which can be Fourier expanded by $e^{im\phi}$ for all $m\in\mathbb{Z}$. In that case, the asymptotic ($r\to\infty$) killing vector takes the following form
\begin{align}\label{bichhagol}
     \xi=\Big(X(\phi)-O\big(r^{-1}\big)\Big)\partial_\phi+\Big(Y(\phi)+u X'(\phi)\Big)\partial_u-\Big(rX'(\phi)-O(r^0)\Big)\partial_r.
\end{align}
Fourier expanding the functions $X$ and $Y$, it is straightforward to show that the symmetry generators at null infinity is
\begin{align}\label{jiraforing}
    L_n=ie^{in\phi}(\partial_\phi+inu\partial_u-inr\partial_r),\qquad M_n=ie^{in\phi}~\partial_u.
\end{align}
These generators satisfy the CCA$_2$ (or BMS$_3$) algebra given in \eqref{BMS}. This establishes CCA$_2$ is indeed the asymptotic symmetry algebra at the null infinity of flat spacetime (or in general, any AFS is CCA$_2$). 
\medskip

Now, we look into the flat spacetime restricted by an EOW brane located at $x^2=r\sin{\phi}=0$. $r\sin{\phi}=0$ for all $r$ means $\sin{\phi}=0$ i.e. either $\phi=0$ ($x^1>0$  region of the $x^2=0$ plane) or $\phi=\pi$ ($x^1<0$ region of the $x^2=0$ plane). Hence, for such boundary, only the generators having no $\partial_\phi$ term at \{$\phi=0,\pi$\} are allowed to be symmetry generators.
\medskip

Let us look again into the symmetry generators of flat spacetime at $r\to\infty$ we found in \eqref{jiraforing}. Introducing $\sin{\phi}=0$ boundary in the bulk implies that it has the following impact on the set of generators $L_n$s
\begin{align}
\begin{split}
   \mathcal{Q}_n&=L_{n}+L_{-n}=2i\cos{(n\phi)}~\partial_{\phi}-2inu\sin{(n\phi)}~\partial_{u}+2inr\sin{(n\phi)}~\partial_{r}~,\\
   \mathcal{O}_n&=L_{n}-L_{-n}=-2\sin(n\phi)~\partial_\phi-2nu\cos(n\phi)~\partial_u+2nr\cos(n\phi)~\partial_r~.
   \end{split}
\end{align}
We can see from the above expressions that, at \{$\phi=0,\pi$\} boundaries, the generators $\mathcal{Q}_n$ have non-zero $\partial_\phi$ terms, rendering them boundary incompatible. However the $\partial_\phi$ term for $\mathcal{O}_n$ vanish at \{$\phi=0,\pi$\} making them boundary compatible.
\medskip

In the set of generators $M_n$s given in \eqref{jiraforing}, there are no $\partial_\phi$ term at the first place which seems to imply that all the generators are boundary compatible. However, 
if we 
choose to work with the Fourier expansion of the killing vector, including subleading terms in \eqref{girgitia}, 
then at large $r$, we find a subleading term involving $\partial_\phi$
\begin{align}
    M_n=ie^{in\phi}\partial_u+\frac{m}{r}e^{im\phi}\partial_\phi+ \cdots
\end{align}
Taking this subleading term into consideration, we have the following expressions for $\mathcal{P}_n$s and $\mathcal{R}_n$s 
    \begin{align}\label{PRSeggregate}
    \begin{split}
\mathcal{P}_n&=M_n+M_{-n}=2i\cos{(n\phi)}\partial_u+\frac{2in}{r}\sin{(n\phi)}\partial_\phi+\cdots\\
\mathcal{R}_n&=M_n-M_{-n}=-2\sin{(n\phi)}\partial_u+\frac{2n}{r}\cos{(n\phi)}\partial_\phi+ \cdots
\end{split}
\end{align}
Now, let us make our demand a little stronger; any symmetry generator, which allows even a tiny fluctuation of order $O(r^{-1})$ on the boundaries $\phi=0,\pi$, will be considered boundary incompatible. Now, one can see in \eqref{PRSeggregate} that the generators $\mathcal{P}_n$s do not have any $\p_\phi$ term at $\phi=0,\pi$ even at order $O(r^{-1})$, while $\mathcal{R}_n$s do have such terms at $O(r^{-1})$
\begin{align}
    \mathcal{R}_n\big|_{\phi=0,\pi}=\frac{2n}{r}\cos{(n\phi)}\p_\phi\big|_{\phi=0,\pi}
\end{align}
In this regard it is worth recalling from section \ref{section2} that while taking Carroll limit of relativistic BCFT generators, we found that the symmetry generators $\mathcal{R}_n$s distort boundaries at subleading order (see \eqref{khurorkal}). These results indicate that $\mathcal{R}_n$s, in general, have a tendency to allow tiny fluctuations. 
\medskip

The generators $\mathcal{P}_n$s at $r\to \infty$ limit takes the following form
\begin{align}
    \mathcal{P}_n=2i\cos{(n\phi)}\partial_u.
\end{align}
Hence, we end up with the following generators    \begin{align}\label{badurgopal}
    \mathcal{O}_n=-2\sin(n\phi)\,\partial_\phi-2nu\cos(n\phi)\,\partial_u+2nr\cos(n\phi)\,\partial_r,\qquad
        \mathcal{P}_n=2i\cos{(n\phi)}\partial_u.
    \end{align}
These generators follow the version of BCCA$_2$ given in \eqref{pro952}, confirming that asymptotic symmetry algebra of flat spacetime (or in general AFS) restricted by $x^2=0$ plane indeed is same as \eqref{pro952}.

\subsection{Flat limit from AdS$_3$}
We now present the derivation of BCCA$_2$ by taking the flat-space limit of the asymptotic symmetries of AdS$_3$ in the presence of an EOW brane. As a preliminary step, we first review the asymptotic symmetry algebra of AdS$_3$ without an EOW brane.

\subsection*{AdS$_3$ without EOW brane}
We begin by writing the asymptotic killing vectors of asymptotically AdS$_3$ spacetime with Brown-Henneaux boundary conditions in terms of ($u,r,\phi$) coordinates \eqref{ucoordinate} 
\begin{align}\label{BHVirasoro}
    \mathcal{L}^\pm_n=ie^{inx^\pm}\Big(\p_\pm-\frac{i}{2}nr\p_r\Big), \qquad     x^\pm=\frac{u}{\ell}\pm\phi.
\end{align}
These generators famously satisfy Virasoro algebra and the central charges are $c^\pm=\frac{3l}{2G}$ \cite{Brown:1986nw}.  
In flatspace ($\ell\to\infty$) limit we take the   In\"on\"u-Wigner contraction \eqref{oldcon} ($\e$ replaced by $\frac{1}{\ell}$) on these generators and retrieve the generators
\eqref{jiraforing}. The central charges in this limit evolve as
\begin{align}
    c_L=c^+-c^-=0,\qquad c_M=\frac{1}{\ell}(c^++c^-)=\frac{3}{G}.
\end{align}
To understand the flat limit ($\ell \to \infty$) physically, we note that
\eqref{ucoordinate} implies
\begin{align}\label{I+toband}
    u=t-\frac{\ell\pi}{2}=\t\ell-\frac{\ell\pi}{2},\qquad  \implies \t=\frac{\pi}{2}+\frac{u}{\ell}.
\end{align}
\begin{figure}[t]
\centering
    \includegraphics[scale=0.8]{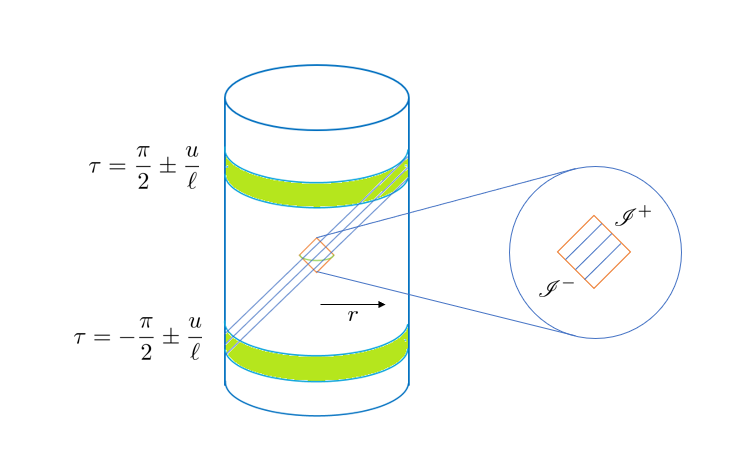}
    \caption{AdS$_3$ spacetime with Minkowski diamond inside. The green band gets mapped to the null infinity of the Minkowski diamond.}
    \label{green band}
\end{figure}
In the above, we have rescaled the timelike coordinate in AdS$_3$ as $\tau=\frac{t}{l}$.  Equation \eqref{I+toband} points toward the fact that in the flat-space limit, only a small band on the AdS$_3$ asymptotic cylinder around $\t=\pm\frac{\pi}{2}$ gets mapped to $\mathcal{I}^\pm$\footnote{In $\mathcal{I}^-$, the null coordinate is advanced time coordinate $v=t+r$ and the corresponding coordinate in AdS$_3$ spacetime is $v=t+\ell\tan^{-1}\frac{r}{\ell}$. At $r\to\infty$ this leads to $v=\t l-\frac{\ell\pi}{2}, \implies \t=\frac{\pi}{2}+\frac{v}{\ell}$.} (as illustrated in Fig \ref{green band}). Consequently, Carrollian correlators come from $\ell\to\infty$ limit of the AdS Witten diagrams if the CFT operators are inserted in the small bands $\t=\frac{\pi}{2}\pm\frac{u}{\ell}$ \cite{Bagchi:2023fbj}.

\subsection*{AdS$_3$ with EOW brane}
Now, let us consider the case of AdS$_3$ spacetime with EOW brane discussed in section \eqref{Ads3Eow}. Since the EOW brane intersects with the AdS$_3$ boundary at $\phi=0,\pi$, we need to consider the combination of generators which are free from $\p_\phi$ terms at $\phi=0,\pi$, and the following generators satisfy that condition
\begin{align}
    \mathbb{L}_n=\mathcal{L}^+_n+\mathcal{L}^-_n=ie^{in\frac{u}{\ell}}~\left(l\cos{n\phi}~\p_\t+i\sin{n\phi}~\p_\phi-inr\cos{n\phi}~\p_r\right).
\end{align}
We apply $\ell\to\infty$ limit on these generators through the following contraction
\begin{align}
    \mathcal{O}_n=\mathbb{L}_n-\mathbb{L}_{-n},\qquad \mathcal{P}_n=\frac{1}{\ell}(\mathbb{L}_n+\mathbb{L}_{-n}).
\end{align}
This contraction leads us to the boundary generators $\mathcal{O}_n$ and $\mathcal{P}_n$ that were found earlier in \eqref{badurgopal}.
The central charge of the BCFT dual of gravity in AdS$_3$ having EOW brane is $c=\frac{3l}{2G}$. Applying this contraction on the $[\mathcal{O},\mathcal{P}]$ commutators \eqref{pro952} one can see that the central charge for the BCCFT appearing in \eqref{pro952} comes from the following flat-space limit of BCFT central charge
\begin{figure}[t]
\centering
    \includegraphics[scale=0.75]{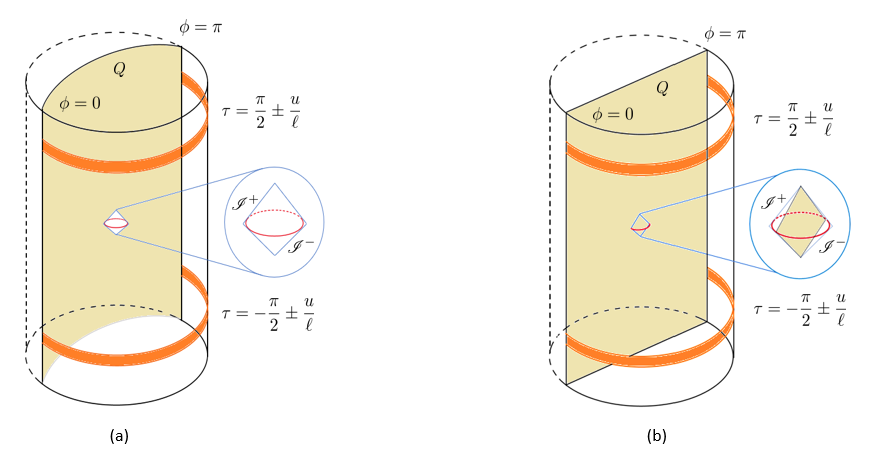}
    \caption{(a) depicts AdS$_3$ spacetime with tensile EOW brane and (b) represents tensionless EOW branes. The tensile brane doesn't intersect the Minkowski diamond, while the tensionless brane does.}
    \label{EOWbranefig}
\end{figure}
\begin{align}
    c_M=\frac{2c}{\ell}=\frac{3}{G}.
\end{align}
This serves as a consistency check for the results derived earlier from intrinsic viewpoint. This symmetry analysis provides us with initial hints of AFS/BCCFT correspondence. 
\medskip

In the light of the fact that in the $\ell\to\infty$ limit, only a small band of AdS boundary cylinder given by $\t=\frac{\pi}{2}\pm\frac{u}{\ell}$ map to the null infinity of the Minkowski diamond, it is expected that here too, the BCCA correlators should come from $\ell\to\infty$ limit of AdS Witten diagrams where the operators are inserted in the small band as depicted in Fig. \ref{EOWbranefig}. We leave this for future work.

\section{Other possible versions of AFS/BCCFT Correspondence}\label{section6}
Holographic duals of BCFTs with different kinds of boundaries have been extensively studied in literature, where each kind boundary corresponds to different kind of EOW branes \cite{Takayanagi:2011zk, Fujita:2011fp,Kawamoto:2022etl,Kusuki:2022ozk}. However, all of these BCFTs may not necessarily have Carrollian counterparts. For example, it is unlikely that 2d BCFT defined on a disk region would have Carrollian counterpart since the notion of circle doesn't make any sense on a 2d Carrollian/null hypersurface\footnote{Geometrically circle is a shape where every point on its boundary are equidistant from a fixed point. On a 2d Carrollian/null hypersurface, distance is not defined along one of the two dimensions, making it difficult to define a Carrollian analogue of disc.}. So far in this work we have discussed possible flat-space/Carrollian analogue of AdS$_3$/BCFT$_2$ on an infinite strip. In this section we briefly discuss about another possible version of flat space holography with EOW brane.
\begin{figure}[t]
\centering
    \includegraphics[scale=0.4]{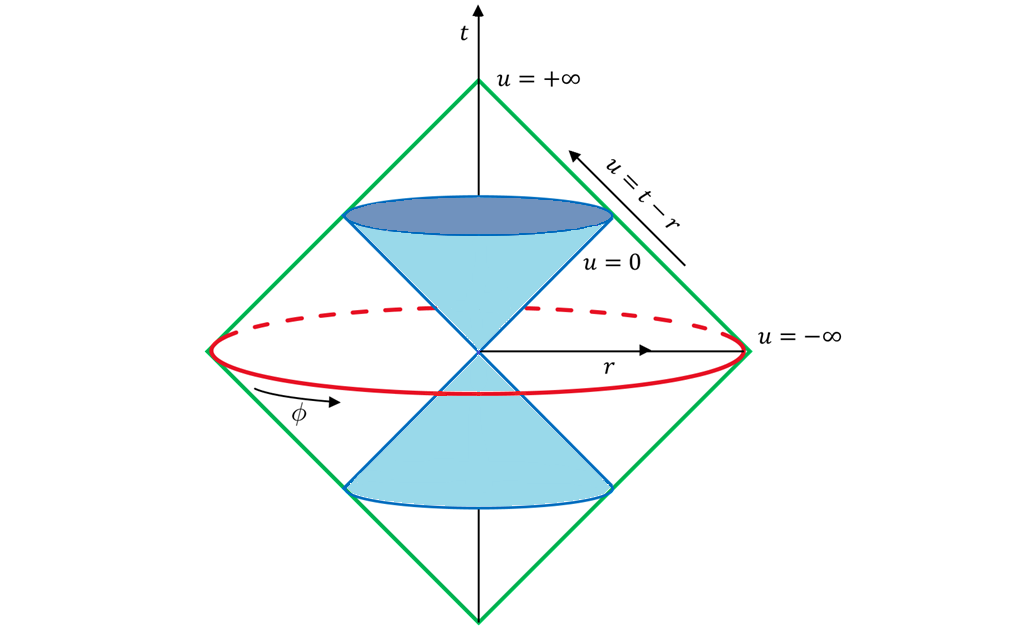}
    \caption{EOW brane coinciding with the light-cone $t=r$}
    \label{nullcyl}
\end{figure}
\subsection*{BCCA with temporal boundary}\label{hukomukhohangla}
Once we introduce $u=0$ boundary\footnote{Here we call $u$ as temporal coordinate and fixed $u$ boundary as temporal boundary.} on a Carrollian cylinder, there will be no $\p_u$ term in the symmetry generators at $u=0$. Looking at the cylindrical CCA$_2$ generators in \eqref{BMSCyl} one can see that $L_n$s have no $\p_u$ term at $u=0$, implying that the entire set of $L_n$s survive the $u=0$ boundary. However $M_n$s have to be discarded completely due to the presence of $\p_u$ regardless of any value of $u$. Hence the surviving generators in this case are given by
\begin{align}\label{timecylcarr}
 L_n =i\,e^{in\phi}(\partial_\phi+inu\,\partial_u).   
\end{align}
Interestingly, the surviving generators give us one copy of Virasoro algebra. Hence, BCCFT$_2$ defined on a null manifold having temporal boundary has to be constructed on one copy of Virasoro algebra.

\subsection*{Bulk realisation}
Let us now consider $u=0$ boundary on $\mathcal{I}^+$. Since $u=t-r$, it is natural to expect the hypersurface $t=r$ as the location of EOW brane which intersects with $\mathcal{I}^+$ at $u=0$. It is obvious that the EOW brane coincides with the lightcone originated at $r=0$ and its intersection with the $\mathcal{I}^+$ gives a future celestial circle (circle spanned by $\phi$ on $\mathcal{I}^+$ for fixed $u$) as shown in Fig. \ref{nullcyl}.  
\medskip

Let us look into the impact of the $t=r$ boundary on this. We first recall the asymptotic Killing vector of flat spacetime without boundaries in \eqref{bichhagol}
\begin{align*}
     \xi=\Big(X(\phi)-O\big(r^{-1}\big)\Big)\partial_\phi+\Big(Y(\phi)+u X'(\phi)\Big)\partial_u-\Big(rX'(\phi)-O(r^0)\Big)\partial_r.
\end{align*}
From here one can see that the supertranslations which come by setting $X(\phi)=0$, $Y(\phi)=e^{in\phi}$ has $\partial_u$ for all values of $u$, making them incompatible with $u=0$ (or $t=r$) brane location at the bulk. However, for $Y(\phi)=0$ and $X(\phi)=e^{in\phi}$, the set of superrotation generators we have is
\begin{align}\label{gechhodada}
 L_n=ie^{in\phi}(\partial_\phi+inu\partial_u-inr\partial_r).   
\end{align}
In these generators there is no $\partial_u$ term at $u=0$ which confirms their compatibility with $u=0$ or $t=r$ brane location at the bulk. These $L_n$s satisfy one copy of Virasoro algebra, which shows that flat spacetime (or AFS) restricted with null EOW brane has one copy of Virasoro algebra as its asymptotic symmetry algebra.

\section{Conclusion}\label{section7}
In this paper, we have taken a first step towards constructing a holographic dual for 2d Carrollian BCFTs. We identified an appropriate set of end-of-the-world (EOW) branes intersecting $\mathcal{I}^+$ at $\phi=0,\pi$ and found the branes to be tensionless. The AdS counterpart of this setup was found to be AdS$_3$/BCFT$_2$ with BCFT$_2$ on a strip. The intrinsically derived properties of EOW branes turned out to be consistent with the flatspace limit of EOW brane introducced in AdS$_3$. We showed that the presence of EOW brane in 3d flat spacetime breaks $\mathfrak{iso}(2,1)$ symmetry to $\mathfrak{iso}(1,1)$ symmetry, algebra of which is isomorphic to the global part of BCCA. This symmetry beaking mirrors the $\mathfrak{so}(2,2)\to\mathfrak{so}(2,1)$ symmetry breaking in AdS$_3$. 
\medskip

We further constructed the representation of $\mathfrak{iso}(1,1)$ from the induced representation of $\mathfrak{iso}(2,1)$ and found that this coincides with the Carroll limit of highest-weight representation of $\mathfrak{so}(2,1)$. We also showed that the full BCCA emerges as the asymptotic symmetry algebra at the null infinity of flat spacetime restricted by this EOW brane. Finally, we showed that, unlike in the relativistic spacetime, in the null/Carrollian spacetime, the choice of boundary alters the surviving symmetry algebra, resulting into different kinds of BCCA. In the bulk side this implies, different kinds of EOW branes in flat spacetime can lead to different kinds of asymptotic symmetries on the null infinity. As an example, a temporal boundary $u=0$ reduces the symmetry to a single copy of Virasoro, which can be realised in the bulk by an EOW brane coinciding with the lightcone $t=r$. This indicates that there are different versions of AFS$_3$/BCCFT$_2$ correspondence constructed on different symmetry algebras.

\subsection*{Future directions}
In this work, we initiated the study of construction of holographic dual of Carrollian CFTs with boundaries. This work engender research in numerous directions some of which we list below.

\begin{itemize}
\item \textit{BCCFT with a boundary along a null direction:} An immediate direction for future work could be to construct the holographic dual of BCCFT with its boundary imposed along a null direction. As briefly discussed in Section~\ref{section6}, such a boundary is realized holographically by an EOW brane described by a light-cone hypersurface, for which the asymptotic symmetry algebra reduces to a single copy of the Virasoro algebra. It would be interesting to develop this correspondence in detail and identify the associated holographic dictionary.

\item \textit{Concrete formulation of Carrollian BCFT:} As highlighted before, the present work constitutes an initial step towards formulation of holography of AFS$_3$ with EOW branes. In order to construct a complete formulation of holographic duality we need to have a concrete understanding of Carrollian BCFT governed by BCCA$_2$\footnote{Here, by BCCA$_2$ we mean the algebra associated with $\phi=0,\pi$ boundaries,  as given in \eqref{pro952}. This should not be confused with the algebra associated with a boundary along a null direction, which gives rise to a single copy of the Virasoro algebra.}. In section \ref{repglobal}, we briefly discussed the possible representation theory of the global subalgebra of BCCA$_2$. A detailed analysis of the representation of full BCCA$_2$ will be presented in an upcoming work \cite{upcomingpaper}. Based on the representation theory, an important next step is to develop the corresponding Carrollian BCFT, including its correlation function and other observables. In case of AdS/BCFT correspondence, geodesic Witten diagrams have been used to understand bulk duals of corelation functions \cite{Karch:2017wgy}.  Analogous study is needed to be performed for flat spacetime in order to make this duality concrete.

\item \textit{Boundary entropy and g-function:}~ A rigorous formalism of BCFT involves boundary states which play a crucial role in understanding open strings and D-branes. They are also used to calculate boundary entropy or $g$-function. The g-function can also be obtained by calculating entanglement entropy. It has been shown in \cite{Takayanagi:2011zk,Fujita:2011fp} that g-function calculated for BCFT exactly matches the same obtained from calculating holographic entanglement entropy. In order to formulate flatspace analogue of this duality we need to formulate Carrollian analogue of boundary states and boundary entropy. Entanglement entropy for Carrollian CFTs and flat-space analogue of Ryu-Takayanagi conjecture \cite{Ryu:2006bv} has been studied in \cite{Bagchi:2014iea, Jiang:2017ecm}. These aspects are also needed to be understood for Carrollian BCFTs.
\end{itemize}

We hope to return to these and other research directions in near future. 
\bigskip

\subsection*{Acknowledgements}
The authors are deeply indebted to Arjun Bagchi for his careful reading of the manuscript and critical comments that led to several improvements. We are grateful to Shankhadeep Chakrabortty for collaboration on related work and to Daniel Grumiller, Tadashi Takayanagi and Hossein Yavartanoo for valuable discussions concerning this work. We also thank Andrew Strominger, Ashoke Sen, Shiraz Minwalla, Jelle Hartong, Stefan Fredenhagen, Wei Song, Akashdeep Roy, Sharang R. Iyer, Anton Pribytok and David Skinner for useful comments and discussions. 
\medskip

PC acknowledges the support of the Indian Institute of Technology Ropar (IIT Ropar) for the institute fellowship. RC is supported by the Mathematical Physics group of Beijing Institute of Mathematical Sciences and Applications (BIMSA). PP is supported by Raychaudhuri Fellowship from Tata Institute of Fundamental Research (TIFR). PP would also like to thank the Simons Foundation for its financial support to attend ``Celestial Holography Annual Meeting 2026", IISER Pune for its hospitality during ``Advances in Black hole Theory workshop 2026" and the participants of both meetings, where this work was presented, for valuable discussions.
\medskip

The authors gratefully acknowledge the people of India for their generous support towards research in basic science.
\newpage

\bibliographystyle{JHEP}
\bibliography{References}

\end{document}